\documentclass[twocolumn,showpacs,preprintnumbers,amsmath,amssymb]{revtex4}

\usepackage{graphicx}
\usepackage{dcolumn}
\usepackage{bm}

\usepackage{longtable}

\usepackage{epsfig}
\usepackage{xspace}
\begin{document}

\title{Inflation without inflatons}

\author{Reuven Opher}
\email{opher@astro.iag.usp.br}
\author{Ana Pelinson}
\email{anapel@astro.iag.usp.br} \affiliation{IAG, Universidade de
S\~{a}o Paulo, Rua do Mat\~{a}o, 1226 \\
Cidade Universit\'aria, CEP 05508-900. S\~{a}o Paulo, S.P.,
Brazil}

\date{\today}

\begin{abstract}
We present a model which predicts inflation without the presence of
inflaton fields, based on the $\epsilon R^2$  and Starobinsky
models. It links the above models to the reheating epoch with
conformally coupled massive particles created at the end of
inflation. In the original Starobinsky model, the reheating era was
created by massless non-conformally coupled particles. We assume
here that non-conformal coupling to gravitation does not exist. In
the $\epsilon R^2$ model, inflation is produced by the gravitational
Lagrangian to which a term $\epsilon R^2$ is added, where $\epsilon$
is a constant and $R$ is the Ricci scalar. Inflation is created by
vacuum fluctuations in the Starobinsky model. Both models have the
same late-inflation time-dependence, which is described by a
characteristic mass $M$. There is a free parameter $H_0$ on the
order of the Planck mass $M_{Pl}$ that determines the Hubble
parameter near the Planck epoch and which depends upon the number
and type of particles creating the vacuum fluctuations in the
Starobinsky model. In our model, we assume the existence of
particles with a mass $m$, on the order of $M$, conformally coupled
to gravity, that have a long decay time. Taking  $m\equiv FM$, we
investigate values of $F=0.5$ and $0.3$. These particles, produced
$\sim 60\,e$-folds before the end of inflation, created the nearly
scale invariant scalar density fluctuations which are observed.
Gravitational waves (tensor fluctuations) were also produced at this
epoch. At $t_{\rm end}$, the Hubble parameter begins to oscillate
rapidly, gravitationally producing the bulk of the $m$ particles,
which we identify as the origin of the matter content of the
Universe today. The time required for the Universe to dissipate its
vacuum energy into $m$ particles is found to be $t_{\rm dis} \simeq
6\,M_{Pl}^2/M^3F$. We assume that the reheating time $t_{RH}$ needed
for the $m$ particles to decay into relativistic particles, is very
much greater than that necessary to create the $m$ particles,
$t_{\rm dis}$. A particle physics theory of $m$ can, in principle,
predict their decay rate
 $\Gamma_{mr}\equiv t_{RH}^{-1}$. From the ratio $f\equiv t_{\rm dis}/t_{RH}$, $F$ and
$g_{\ast}$ (the total number of degrees of freedom of the
relativistic particles) we can, then, evaluate the maximum
temperature of the Universe $T_{\rm max}$ and the reheat temperature
$T_{RH}$ at $t_{RH}$. From the observed scalar fluctuations at large
scales, $\delta\rho/\rho\sim 10^{-5}$, we have the prediction
$M\cong 1.15\times 10^{-6} M_{\rm Pl}$ and the ratio of the tensor
to scalar fluctuations, $r\cong 6.8\times 10^{-4}$. Thus our model
predicts $M$, $t_{\rm dis}$, $t_{\rm end}$, $T_{\rm max}$, $T_{RH}$,
$t_{\rm max}$, and $t_{RH}$ as a function of $f$, $F$, and
$g_{\ast}$ (and to a weaker extent the particle content of the
vacuum near the Planck epoch). A measured value of $r$ that is
appreciably different from $r=6.8\times 10^{-4}$ would discard our
model (as well as the Starobinsky and $\epsilon R^2$ models).

\end{abstract}

\pacs{$\,\,$ 98.80.-k,$\,\,$   98.80.Es}
\maketitle

\section{Introduction}

The standard model of inflation, based on the existence of a scalar
inflaton field, makes the following assumptions:

\begin{enumerate}
  \item The beginning of inflation occurs at an energy $<<
M_{Pl}$ (Planck energy). Its origin is unknown and the state of the
Universe before the beginning of inflation is undefined;
  \item A large initial displacement of $\phi$
from the minimum of $V(\phi)$ is necessary for the onset of
inflation (such as in the chaotic inflation model);
  \item The potential energy of the
inflaton dominates its kinetic energy during inflation; and
  \item The inflaton potential, $V(\phi)$, and its first derivative are defined by
observations $\sim 60\,e$-folds before the end of inflation.
\end{enumerate}
In complex inflation theories, there can be more than one inflaton.
(See \cite{basset05} for a recent review of inflation theory with
inflatons.)

Here we present a model which links the Starobinsky and $\epsilon
R^2$ models of inflation, where $R$ is the Ricci scalar to the
reheating era. All three models, the Starobinsky, $\epsilon R^2$
and ours, do not involve inflatons to create inflation. They also
avoid most of the above assumptions.

In the Starobinsky model \cite{star1}, an $R^2$ term in the
effective Lagrangian dominates inflation at late times (see also
\cite{suen}, \cite{mijic}). There is no sharp boundary between the
Starobinsky model and $\epsilon R^2$ model since the latter is the
particular case of the former in the limit $M \ll H_0$ (using the
notation of Eq.(9), with some small non-local terms (due to non-zero
rest masses of conformally coupled quantum fields) omitted. However,
the Starobinsky and $\epsilon R^2$ models have the same qualitative
behavior at $\sim 60\,e$-folds before the end of inflation, when the
presently observed scalar and tensor fluctuations were produced.

Both the Starobinsky and $\epsilon R^2$ cosmologies are
characterized by a single mass $M$ ($\equiv
M_{Pl}/\sqrt{24\,\epsilon}$ in the $\epsilon R^2$ cosmology and
$M_{Pl}/\sqrt{48\pi k_1}$ in the Starobinsky model, where
$M_{Pl}\equiv G^{-1/2}$ is the Planck mass and $k_1$ is the
coefficient of the term which contains the second derivative of $R$
in the quantum corrected vacuum expectation value of the
energy-momentum tensor [Eq.(\ref{anomaly})]). The mass $M$
characterizes the end of the inflation period, during which, the
Hubble parameter varies slowly. A period then begins, in which $H$
oscillates rapidly as $H\propto (1/t)\cos^2{\omega t}$ and the
cosmological scale factor varies as $a(t)\propto
t^{2/3}[1+\sin{[2\omega t]}/(3\omega t)]$, where $\omega \simeq
M/2$. When averaged over several oscillations, the Universe expands
as a classical matter-dominated Universe.

Although the $\epsilon R^2$ model can be considered to be the
simplest way to produce inflation, i.e., by means of a simple
modification of the gravitational Lagrangian, we concentrate here on
the Starobinsky model since it is more complete. It links the
beginning of the inflation period to the beginning of the Universe
and also describes the end of inflation in detail.

The Starobinsky model suggests that for energy densities and
curvatures near the Planck scale, quantum corrections to Einstein's
equations become important (as discussed in detail by Vilenkin
\cite{vile85a}). In the Starobinsky model, inflation is driven by
one-loop corrections due to quantized matter fields \cite{star1}
(see also \cite{star2,fhh,Mukhanov81,ander,anapel1,anapel2}). The
model is consistent with a Universe that was spontaneously created,
as discussed by Grishchuk and Zel'dovich \cite{grish1}.

The beginning of the Starobinsky inflation period can be associated
with the beginning of the Universe due to quantum fluctuations of
the vacuum. Tryon \cite{tryon} was the first to suggest that a
closed Universe can be created spontaneously as a result of a
quantum fluctuation. Vilenkin \cite{vile85a,vile8234,vile85b},
Zel'dovich and Starobinsky \cite{zel1}, and Linde \cite{linde} were
the first to attempt to describe the quantum creation of a Universe
in the framework of quantum gravity. The picture that emerges is one
of a Universe tunneling quantum mechanically to a de Sitter space
time. At the moment of nucleation ($t=0$), the Universe has a size
$a(0)=H_{\rm in}^{-1}$. This is the beginning of time and, from that
point on, the Universe evolves along the lines of the inflation
scenario.

In the Starobinsky model, inflation is produced by the vacuum energy
$\rho_V$, which has negative pressure, $P = -\rho_V$. Inflation in
both the Starobinsky and our models can be described by an effective
geometric scalar particle $M$. In our model, there is an additional
massive particle $m$ produced at the end of inflation, which is
freely moving and which produces positive pressure.

Structure in the Universe primarily comes from almost
scale-invariant superhorizon curvature perturbations
\cite{giudice01,giudice04}. In our model, a mass $m$ is much less
than the Hubble parameter during inflation. The mechanism of $m$
particle production from inflation is based on the observation that
particles that are massive in the present-day vacuum, could have
been very light during inflation. This implies that fluctuations of
a generic scalar field $\chi$ with mass $m \ll H$ during inflation
are copiously generated, with an almost scale invariant spectrum
\cite{linde90,liddle,riotto}. The particles become heavy and
non-relativistic at the end of inflation.

The end of the Starobinsky inflation period has been suggested to be
due to the masses of the particles in the vacuum fluctuations [14].
Thus, since the mass $M$ describes the end of inflation in both the
Starobinsky and our models, $M$ is a natural mass scale for the
particles that are created at the end of inflation. In our model, we
then have the scenario that particles of mass comparable to the mass
$M$ in the vacuum fluctuations first create the inflation, after
which, particles of mass $m$ comparable to, but slightly less than
$M$, are produced from the vacuum due to the rapid change of the
Hubble parameter. The particles $m$ are conformally coupled to
gravitation (Ricci scalar). These massive particles create the
reheating epoch of the Universe.

Our model can be compared with that of the Starobinsky model, in
which massless particles, non-conformally coupled to gravitation,
directly create the reheating era. Here we assume that non-conformal
coupling does not exist.

Previously, gravitational production of massive particles has been
investigated in order to explain the observed ultra-high energy
cosmic rays, produced as a result of heavy particle decay (masses
$>10^{12}\,{\rm GeV})$ \cite{cosmray}. The gravitational particle
production of the heavy particles  $m$ can be described assuming a
given background metric \cite{kuzmin,kolb}.

The paper is organized as follows. In section II, we give the main
results of the Starobinsky model, as discussed by Vilenkin
\cite{vile85a}. We discuss the gravitational production of the $m$
particles in section III. In section IV, we derive the scalar
density fluctuations produced $\sim 60\,e$-folds before the end of
inflation. We obtain the ratio $r$ of the tensor to scalar
fluctuations in section V. The reheating of the Universe is
discussed in section VI. Our conclusions and discussion are
presented in section VII.

\section{The Starobinsky inflationary model}

In this section, we discuss the Starobinsky model of inflation,
following the description and notation of Vilenkin. This model is
based on the semiclassical Einstein equations,
\begin{equation} R_{\mu\nu}-\frac{1}{2} g_{\mu\nu} R=-8\pi G
<T_{\mu\nu}>\,,
\end{equation}
which assume a spatially flat Robertson-Walker metric,
\begin{equation} ds^2= a^2(\eta) ( d\eta^2 - d{\bf x}^2)\,,
\label{coord}
\end{equation}
where $dt=a d\eta$, $t(\eta)$ is the proper (conformal) time and
$a$ is the scale factor.

The quantum corrections to the expectation value of the
energy-momentum tensor for massless particles in curved space time
are
\begin{equation} <T_{\mu\nu}> = k_1\,^{(1)}H_{\mu\nu}+ k_3\,^{(3)}H_{\mu\nu}\,,
\label{anomaly}
\end{equation}
where
$$ ^{(1)}H_{\mu\nu}=2\,R_{;\,\mu;\,\nu}-2g_{\mu\nu}\,
{R_{;\,\sigma}}^{;\,\sigma} +
2RR_{\mu\nu}-\frac{1}{2}g_{\mu\nu}R^2\,,
$$
\begin{equation} ^{(3)}H_{\mu\nu}={R_{\mu}}^{\sigma}R_{\nu\sigma}-\frac{2}{3}R\,R_{\mu\nu}-
\,\frac{1}{2}g_{\mu\nu}R^{\sigma\tau}R_{\sigma\tau}+
\frac{1}{4}g_{\mu\nu}\,R^2 \,,\label{Htensor}
\end{equation}
and $k_1$, $k_3$ are constants. The value of the coefficient $k_1$
can be determined by observations. However, the value of $k_3$ is
fixed by the condition
\begin{equation} k_3\,=\, \frac{1}{1440\,\pi^2}\,\Big( N_0 +
\frac{11}{2}\,N_{1/2} + {31\,N_1}\Big)\,,
\label{b}
\end{equation}
where $N_0,\,N_{1/2},$ and $N_1$ are the numbers of quantum matter
fields with spins $0,\,1/2,\,$ and $1$, respectively.

It is convenient to define two new parameters, $H_0$ and $M$, in
terms of $k_1$, $k_3$, and the Planck mass, $M_{Pl}$:
\begin{equation}
H_0=\,\frac{M_{Pl}}{\sqrt{8\pi k_3} }\, \label{hubbstar}
\end{equation}
and
\begin{equation} M=\,\frac{M_{Pl}}{\sqrt{48\pi k_1}}\,.
\label{mstarvalue}
\end{equation}
The Planck constant $\hbar$ and the speed of light $c$ are given in
natural units, $\hbar=c=1$, and $M_{Pl}={G}^{-1/2}\cong 1.22 \times
10^{19} {\rm GeV}$. To evaluate $H_0$, a minimal SU(5) model, where
$N_0=34$, $N_{1/2}=45$, $N_1=24$, is assumed, giving
\begin{equation}
H_0\cong 0.74\,M_{Pl}\,.\label{H0su5}
\end{equation}

From the above, the energy-momentum tensor in Eq.(\ref{anomaly})
is
\begin{equation} <T_{\mu\nu}> =\,\frac{1}{8\pi}\left[ \frac{1}{6}\,^{(1)}
H_{\mu\nu}\,\Big(\frac{M_{Pl}}{M} \Big)^2 +\,
^{(3)}H_{\mu\nu}\,\Big(\frac{M_{Pl}}{H_0} \Big)^2\right] \,.
\label{anom}
\end{equation}
Whereas $T^{\nu}_{\nu} =0$ for classical conformally invariant
fields, a trace anomaly arises from the process of regularization
(see \cite{birrel,bunch,dowker}),
\begin{equation} <T^{\nu}_{\nu}>=\,\frac{M_{Pl}^2}{8\pi H_0^2}
\left[\frac{1}{3}\,R^2- R_{\nu\sigma}R^{\nu\sigma}-
\left(\frac{H_0}{M}\right)^2{R_{;\,\nu}}^{;\,\nu}\right]\,.
\end{equation}
The evolution equation for the Hubble parameter $H(t)(= \dot{a}/a)$
in a flat Universe is
\begin{equation} H^2\,(H^2-H^2_0)=\Big(\frac{H_0}{M}\Big)^2\,(2\,H\,
\ddot{H}+6\,H^2\,\dot{H}- {\dot
H}^2)\,.
\label{flat}
\end{equation}
Assuming that $H$ is slowly varying during inflation, $\dot{H}\ll
H^2$, and that $\ddot{H}\ll H\,\dot{H}$, the solution of
Eq.(\ref{flat}) is
\begin{equation} H=H_0\,\tanh{\Big[\gamma-\frac{M^2\,t}{6\,H_0}\Big]}\,,
\label{Hsol}
\end{equation}
where
\begin{equation}
\gamma={1}/{2}\,\ln [{2}/{{\delta}_0}]\,, \label{gdelta}
\end{equation}
\begin{equation}
{\delta}_0=\left|H_{\rm in}-H_0\right|/H_0\label{deltah}\,,
\end{equation}
and $H_{\rm in}$ is the initial value of the Hubble parameter in
the inflation era.

We see from Eq.(\ref{Hsol}) that $H$ changes on a time scale,
\begin{equation} \tau \sim 6\,H_0/M^2\,.\label{tau}
\end{equation}
A long period of inflation occurs when $ M^2\ll 6\,H_0^2\,. $ The
solution of Eq.(\ref{Hsol}) in terms of $a$, valid up to the time at
the end of inflation, when $H\sim M$, is
\begin{equation} t_{\rm end}=\frac{6\,\gamma\,H_0}{M^2}\,,
\label{tend}
\end{equation}
is
\begin{equation}
a(t)=H_0^{-1}\Big(\frac{\cosh{\gamma}}
{\cosh{[\gamma-t/\tau]}}\Big)^{H_0\,\tau}\,\, \label{Hsoltau}\,,
\end{equation}
where $\tau$ is given by Eq.(\ref{tau}).

Assuming that $H\ll H_0$ near the end of inflation,
Eq.(\ref{flat}) simplifies to
\begin{equation}
2 H \ddot{H} + 6 H^2 \dot{H} - {\dot H}^2 + M^2 H^2 = 0\,,
\label{flatmod}
\end{equation}
which has the solution
\begin{equation}
H(t)\cong \frac{4}{3\,t}\, \cos^2{\Big[ \frac{M t}{2}\Big]}\Big(
1-\frac{\sin{[M t]}}{M t}\Big)\, \label{Haprox}
\end{equation}
and
\begin{equation}
a(t)\propto t^{2/3}\Big[ 1+\frac{2}{3 M t}\Big]\,\sin{[M t]}\,.
\label{asolvile}
\end{equation}
Thus, from Eqs.(\ref{Haprox}) and (\ref{asolvile}), $H(t)$ and
$a(t)$ are in an oscillating phase at the end of inflation. At a
time $t_0 \gg M^{-1}$, the period of oscillation is much shorter
than the average Hubble time, $2\,/3\,t_0$. For a time interval
$t_0\gg \Delta t\gg M^{-1}$, we can neglect the power law expansion
in Eq.(\ref{asolvile}), so that
\begin{equation}
a(t)\cong 1+ \Big(\, \frac{2}{3Mt_0} \Big)\sin{[Mt]}\,.
\label{asin}
\end{equation}

In the Starobinsky model, massless particles are gravitationally
produced at the end of inflation. For a scalar field of mass $m$
that satisfies the equation
\begin{equation} {\Box} {\phi}+ (m^2+ \xi R ) \phi=0\,,
\label{phieq} \end{equation}
the field is conformally coupled if $\xi=1/6$ and non-conformally
coupled if $\xi\neq 1/6$. For $m>M/2$, the particle production is
exponentially depressed. In a conformally flat spacetime, massless
conformally coupled particles cannot be gravitationally produced
\cite{star2,grish2}. Therefore, these particles must be
non-conformally coupled.

The oscillation term in Eq.(\ref{asin}) is small and can be
considered to be a perturbation in the calculation of particle
production. A perturbative technique for calculating the
production of very low mass $m$ ($\ll M$) particles was developed
by Zel'dovich and Starobinsky \cite{partprod} and Birrel and
Davies \cite{partprod2}, treating $\left|\,\xi-1/6\right|$ as a
very small parameter. They showed that for a massless
non-conformally coupled field ($m=0,\,\xi \neq 1/6$), the particle
production rate is
\begin{equation}
\dot{n}=\frac{1}{16\pi}\,\Big(\xi-\frac{1}{6} \Big)^2\,R^2\,, \label{nrate}
\end{equation}
where
\begin{equation}
R\approx 6\,\ddot{a}=-\Big(\frac{4M}{t_0} \Big)\sin{[Mt]}\,.
\end{equation}

Taking the average of the particle production over the period of
oscillation and using the fact that the particles are produced in
pairs with energy $m/2$ per particle, the average rate of energy
loss is $(m\equiv F M)$
\begin{equation}
\overline{\dot{\rho}}=\frac{F M\,}{2}\,\overline{\dot{n}}= \frac{F
M^3}{4\pi\,t^2_0} \Big(\xi -\frac{1}{6} \Big)^2\,. \label{rhobardot}
\end{equation}
The rate at which the vacuum energy is dissipated into particles is
\cite{starnonsing}
\begin{equation}
\Gamma\equiv {\overline{\dot{\rho}}}/{\overline{\rho}}= \frac{3}{2}
 \frac{F M^3}{M_{Pl}^2} \Big(\xi -\frac{1}{6} \Big)^2\,\,, \label{gama}
\end{equation}
where
\begin{equation}
\overline{\rho}=\frac{t_0^2}{6\pi G}\,
\label{rhomatt}
\end{equation}
is the energy density of the particles.

Since $ \overline{a}\propto t^{2/3}\,, $ we have
\begin{equation} \overline{H}=
\frac{2}{\,3t_0}\,,
\end{equation}
so that Eq.(\ref{rhomatt}) becomes
\begin{equation}
\overline{\rho}=\frac{3}{8\pi G}\,\overline{H}^{\,2}
\end{equation}
(the Friedmann relation).

\section{Gravitational production of particles}

As discussed in section II and in detail by Vilenkin, massless
particles ($\ll M$) are non-conformally produced at the end of
inflation in the Starobinsky model \cite{star1}. These massless
particles are assumed to reheat the Universe.

Although the main purpose of the analysis of Vilenkin was to
evaluate the gravitational production of massless non-conformally
coupled particles in the Starobinsky model [Eq.(\ref{gama})], an
expression for the conformally coupled gravitational production of
massive $m$ particles was also obtained. We use this expression in
our analysis of massive $m$ particles.

The created $\phi$ field [Eq.(\ref{phieq})] can be expanded in terms
of creation and annihilation operators,
\begin{equation}
\phi(x)=\int{d^3k\left[a_k u_k(x)+a_k^{\dag} u_k^{*}(x) \right] }\,,
\end{equation}
where
\begin{equation}
u_k(x)=\frac{1}{(2\pi)^{3/2}}\, a^{-1}(t)\, e^{i
\overrightarrow{k}\times\overrightarrow{x} }\chi_k(t)
\end{equation}
and the functions $\chi_k(t)$ satisfy the equation for the field
$\chi$ \cite{birrel},
\begin{equation}
\ddot{\chi}_k+k^2 \chi_k+\left[m^2+\left(\xi - \frac{1}{6} \right)R
\right] a^2 \chi_k = 0\,. \label{chieq}
\end{equation}
Linearizing Eq.(\ref{chieq}) and using Eqs.(\ref{asin}) and
(\ref{nrate}), we have
\begin{equation}
\ddot{\chi}_k+{\omega_k}^2 \chi_k-\left[\frac{4}{3 Mt_0} \right]\tilde{m}^2 \sin{[M t]}
\,\,\chi_k = 0\,,\label{lineq}
\end{equation}
where
\begin{equation}
{\omega_k}=(k^2+m^2)^{1/2}\, \label{wkey}
\end{equation}
and
\begin{equation}
\tilde{m}^2=m^2-3 \left(\xi - \frac{1}{6} \right)M^2\,.\label{mrel}
\end{equation}
If the main contribution to the particle production comes from the
modes with $k\sim M/2$ and the mass $m\ll M/2$, we can replace
$\omega_k^2$ with $k^2$ in Eq.(\ref{wkey}). The expression of the
production of $m$ particles for $\xi=1/6$ (conformal production),
\begin{equation}
\Gamma=\frac{G {\tilde m}^4}{6M}  \label{gamaconf}\,
\end{equation}
was presented in the original paper \cite{star1}.

As noted by Vilenkin, although Eq.(\ref{gamaconf}) was derived
assuming $m\ll M$, it also gives a correct order of magnitude for
the conformally coupled decay rate for $m\sim M$, ($m\equiv F M$,
$F=0.3,\,0.5$)
\begin{equation}
\Gamma=\frac{F^4}{6}  \frac{M^3}{M_{Pl}^2} \label{gamaconf2}\,.
\end{equation}

We can compare the time it takes to dissipate the vacuum energy due
to the gravitational production of massless particles,
non-conformally coupled, using Eq.(\ref{gama}),
\begin{equation}t_{\rm dis\xi}(\equiv \Gamma^{-1})=\frac{2}{3}
\frac{M_{Pl}^2}{F^3 M^3} \frac{1}{\left(\xi -{1}/{6}\right)^2}\,,
\end{equation}
with that of massive particles $M$, conformally coupled, from
Eq.(\ref{gamaconf2}),
\begin{equation}t_{\rm dis}=\frac{6}{F^4}
\frac{M_{Pl}^2}{M^3} \,. \label{tdis}
\end{equation}
The ratio of the two times is
\begin{equation}
\frac{t_{\rm dis\xi}}{t_{\rm dis}}=\frac{F}{9 \left(\xi
-{1}/{6}\right)^2}\,. \label{ratio}
\end{equation}
Since it is generally assumed that $\left(\xi -{1}/{6}\right)\ll 1$,
the vacuum energy is dissipated more rapidly in the case of the
emission of the massive $m$ conformally coupled particles
[Eq.(\ref{ratio})].

In general, the vacuum can lose energy, both by the gravitational
production of $m$ particles that are conformally coupled or by
non-conformal gravitational production of massless relativistic
particles. For simplicity, we assume that $\xi=1/6$ and, thus, that
only conformal production of $m$ particles exist. From
Eqs.(\ref{tdis}) and (\ref{Mrrel}) (below), we have
\begin{equation}
{t_{\rm dis}}\simeq \frac{1}{F^4}\,6.8\times 10^{13} \,{r}^{-3/2}\,
t_{Pl}\,,
\end{equation}
where $r$ is the ratio of tensor to scalar fluctuations and
$t_{Pl}=M_{Pl}^{-1}\cong 5.39\times 10^{-44} {\rm sec}$ is the
Planck time.

In this paper, we assume the possible existence of elementary
particles of mass $m\equiv F M$, which are conformally produced
during or at the end of inflation. From their lifetime, we obtain
the maximum temperature of the Universe $T_{\rm max}$ the reheat
temperature $T_{RH}$ and their respective times, $t_{\rm max}$, and
$t_{RH}$. From the gravitational production of the $m$ particles at
$\sim 60\,e$-folds before the end of inflation, we obtain the scalar
density fluctuations (section IV).

\section{Scalar density fluctuations}

Structure in the Universe primarily comes from nearly
scale-invariant superhorizon curvature perturbations. These
perturbations originate from the vacuum fluctuations during the
nearly exponential inflation. In our model, structure is due to a
scalar field $\chi$ of mass $m\ll H_{\rm 60}$, where $H_{\rm 60}$ is
the Hubble parameter $\sim 60\,e$-folds before the end of inflation.
The scalar field $\chi$ of mass $m\ll H_{\rm 60}$ in a quasi-de
Sitter phase was shown to produce quantum fluctuations, whose power
spectrum is scale invariant if they are superhorizon
\cite{linde90,liddle}.

During the inflationary epoch, the fluctuations of the field $\chi$
of mass $m$ obey the equation
\begin{equation}
\delta\ddot{\chi}_k+ 3H\delta\dot{\chi}_k+\left(\frac{k}{a}\right)^2
\delta \chi_k=0\,
\end{equation}
\cite{giudice04}. The fluctuations $\delta \chi$ are described in
terms of the variance,
\begin{equation}
<\chi^2>=\int{\frac{d^3k}{(2\pi)^3}}\,|\delta \chi_k|^2\,,
\end{equation}
which obeys the equation
\begin{equation}
\frac{d<\chi^2>}{dt}=\frac{H^3}{4\pi^2}\,, \label{xi2}
\end{equation}
during inflation \cite{linde90}. (The formula Eq.(\ref{xi2}) was
first independently obtained in \cite{linde82,vile82,star82}.) In
our model,
\begin{equation}
H\cong H_0 \left[\gamma -\frac{M^2 t}{6 H_0}\right]\, \label{Hgamma}
\end{equation}
until the end of inflation, before oscillations set in.
Substituting Eq.(\ref{Hgamma}) into Eq.(\ref{xi2}), we obtain
\begin{equation}
\frac{d <\chi^2>}{dt}=\frac{H_0^3 {\gamma}^3}{4\pi^2} \left[1
-\frac{t}{t_{\rm end}}\right]^3\,, \label{dresult}
\end{equation}
where $t_{\rm end}$ is given by Eq.(\ref{tend}).

We note that the major contribution to the variance comes from
$t\ll t_{\rm end}$, while very little comes from $t\sim t_{\rm
end}$. Integrating Eq.(\ref{dresult}) from $(1/60) t_{\rm end}$ to
$t_{\rm end}$, we estimate the variance of the fluctuations that
were created $\sim 60\, e$-folds before the end of inflation,
\begin{equation}
<\chi^2>\,\simeq \frac{3}{8 \pi^2} \frac{ H_0^4 \gamma^4}{ M^2}
\,. \label{chimedio}
\end{equation}

From Eq.(\ref{Haprox}), the Hubble parameter is
\begin{equation}
H_{\rm end}\simeq \frac{4}{3\,t_{\rm end}}\,
\label{Hend}
\end{equation}
at the end of inflation. Substituting  Eq.(\ref{tend}) into
Eq.(\ref{Hend}), we find
\begin{equation}
H_{\rm end}\simeq \frac{2}{9}\frac{M^2}{\gamma H_0}\,.
\label{Hendgama}
\end{equation}

In a flat Universe, we have
\begin{equation}
H_{\rm end}^2= \frac{8 \pi}{3 M_{Pl}^2}\,[\rho_{V_{\rm end}} +
\rho_{M_{\rm end}}]\,, \label{friedend}
\end{equation}
where $\rho_{V_{\rm end}}$ is the vacuum energy density at the end
of inflation. From Eqs.(\ref{Hendgama}) and (\ref{friedend}),
$\rho_{V_{\rm end}}$ is given by
\begin{equation}
\rho_{V_{\rm end}}=\frac{1}{54\pi} \frac{M^4 M_{Pl}^2 }{\gamma^2
H_0^2}\,,\label{rhoend}
\end{equation}
assuming that $\rho_{V_{\rm end}}$ is very much greater than
$\rho_{M}$ at the end of inflation.

The long wavelength $\chi$ modes satisfy the equation
\begin{equation}
\delta\ddot{\chi}_k+ 3H\delta\dot{\chi}_k+m^2 \delta \chi_k=0\,
\end{equation}
\cite{giudice04}, which gives
\begin{equation}
\delta{\chi}_k \propto a^{-3/2} \cos{m t}\,.
\end{equation}
They are non-relativistic, behaving like a classical homogeneous
field. Their number density is given by
\begin{equation}
n_{m_{\rm end}}=\frac{\rho_{m_{\rm end}}}{m}=\frac{m
<\chi^2>}{2}\,.\label{nchi}
\end{equation}
From Eq.(\ref{chimedio}), we then have
\begin{equation}
\rho_{m_{\rm end}}=\frac{3  H_0^4 \gamma^4 F^2}{16 \pi^2}\,.
\label{rhochi}
\end{equation}

We assume that the $m$ particles decay into relativistic particles
with a decay rate, $\Gamma_{mr} \ll \Gamma_{Vm}=t_{\rm dis}^{-1}$.
Thus, we can separate the decay of the vacuum into $m$ particles
$(\Gamma_{Vm})$ from the decay into relativistic particles.

For the production of the $m$ particles, we have the relation
\begin{equation}
\dot{\rho}_m + 3 H \rho_m =  \dot{\rho_V}\,, \label{vaccons}
\end{equation}
where the second term on the left describes the dilution of the $m$
particles due to the expansion of the Universe and the term on the
right, the production of $m$ particles due to the decay of the
vacuum.

The production of $m$ particles starts at the end of the inflation
period, when
\begin{equation}
\rho_V(t=t_{\rm end})=\rho_{V_{\rm end}}\,
\end{equation}
and
\begin{equation}
\rho_m(t=t_{\rm end})=\rho_{m_{\rm end}} \ll \rho_{V_{\rm end}}\,.
\end{equation}

During the time $m^{-1}\ll t\ll t_0$, when the $m$ particles were
produced, the average Hubble value was $\overline{H}=2/3t_0$. The
vacuum energy decays in a time $t_{\rm dis}=\Gamma_{Vm}^{-1}$,
during which, the $m$ particles are produced:
$\rho_V(t)=\rho_{V_{\rm end}} \exp{[-\Gamma_{Vm} t]}$. Let us take
$t_0\cong t_{\rm dis}$. We can then describe the production of the
$m$ particles by Eq.(\ref{vaccons}) in the form
\begin{equation}
\dot{\rho}_m + 3 \overline{H} \rho_m = \rho_{V_{\rm end}}/t_{\rm
dis}\,. \label{vacdecay}
\end{equation}
The solution of the homogeneous form of Eq.(\ref{vacdecay}) is
\begin{equation}
{\rho}_m (t) = {\rho}_{m_{\rm end}} e^{-3 \overline{H} (t - t_{\rm
end})}
\end{equation}
and that of the inhomogeneous form,
\begin{equation}
{\rho}_m (t) = {\rho}_{m_{\rm end}} e^{-3 \overline{H} (t - t_{\rm
end})} + \frac{\rho_{V_{\rm end}}}{t_{\rm dis}} (t - t_{\rm end})\,.
\label{mdecaysol}
\end{equation}

We take the initial time for the decay of the $m$ particles into
relativistic particles to be
\begin{equation}
t_{mri}\cong t_{\rm dis}\gg t_{\rm end}\,. \label{tmri}
\end{equation}
From Eq.(\ref{mdecaysol}), we have
\begin{equation}
{\rho}_m (t=t_{mri})\equiv {\rho}_{mi}  \simeq \rho_{V_{\rm end}}\,.
\label{rhomi}
\end{equation}

Since the Universe had an average cosmic scale factor $\overline{a}
\propto t^{2/3}$ from $t_{\rm end}$ to $t_{\rm dis}$, it expanded by
a factor $(t_{\rm dis}/t_{\rm end})^{2/3}$ and diluted $\rho_{m_{\rm
end}}$ by a factor $(t_{\rm end}/t_{\rm dis})^{2}$ during this time
interval. Observations show that the ratio of the energy density of
the mass fluctuations at $t_{\rm dis}$, $\rho_{m_{\rm end}} (t_{\rm
end}/t_{\rm dis})^2$, to that of matter, $\rho_{mi}$, is $\sim
10^{-5}$.

From Mukhanov and Chibisov (1981) \cite{Mukhanov81} we have an
approximately flat scalar fluctuation spectrum for the Starobinsky
model with an amplitude
\begin{equation}
\delta_{ST}\sim 3\left(\frac{M}{M_{\rm
Pl}}\right)\left(\frac{8\pi}{3}\right)^{1/2}\,.\label{starfluct}
\end{equation}
From Eqs.(62) and (63), we then have
\begin{equation}
\frac{(t_{\rm end}/t_{\rm dis})^2\,\rho_{\rm mend}}{\rho_{\rm
Vend}}\simeq 10^{-5}\simeq \delta_{ST}\,. \label{fracrhobs}
\end{equation}
Although $\delta_{ST}$ was evaluated for massless particles,
similarly to Vilenkin [8], as discussed in Sec.II, we assume that
this is also the amplitude for the production of massive particles
on the order of $M$. From Eqs.(64) and (65),
\begin{equation}
{M}\simeq 1.15\times 10^{-6} M_{\rm Pl}\,. \label{mstar}
\end{equation}

\section{Ratio of tensor to scalar fluctuations}

From Vilenkin, the tensor power spectrum is
\begin{equation}
P_T(k)=\left\vert { h_k }\right\vert^2=\frac{G
M^2}{2\pi^2}\frac{1}{k^3}
\end{equation}
and the scalar power spectrum is
\begin{equation}
P_S(k)=\delta_k^2=\frac{{\left(\delta\rho/\rho\right)}_{\rm
hor}^2}{k^3}\,,
\end{equation}
where $k$ is the wavenumber of the fluctuations and
${\left(\delta\rho/\rho\right)}_{\rm hor}^2$ are the density
fluctuations on the order of the horizon. The ratio $r$ of tensor to
scalar fluctuations is, then,
\begin{equation}
r\equiv\frac{ P_T(k)}{P_S(k)}=\frac{G
M^2}{2\pi^2}\frac{1}{{\left(\delta\rho/\rho\right)}_{\rm hor}^2}\,.
\end{equation}
We used the observed value of the scalar fluctuations,
${\left(\delta\rho/\rho\right)}_{\rm hor}\simeq 10^{-5}$, to obtain
the value of $M$ in terms of $r$,
\begin{equation} M\simeq 4.4 \times 10^{-5} \sqrt{r}\, M_{Pl}\,.
\label{Mrrel}
\end{equation}

From Eq.(\ref{tdis}) we obtain $t_{\rm dis}$. We find that for
$F=0.5$, $t_{\rm dis}\simeq 6.28 \times 10^{19} t_{Pl}$ and that for
$F=0.3$, $t_{\rm dis}\simeq 4.85 \times 10^{20} t_{Pl}$.

We have
\begin{equation}
r\simeq  6.8\times 10^{-4}\,
\end{equation}
from Eqs.(\ref{Mrrel}) and (\ref{mstar}).

A small value for $r$, bounded from below by $r> 3\times 10^{-6}$
(unless $V^{'''}/V$ in the inflaton potential is unreasonably
large), was previously indicated in \cite{steen}.

From Eqs.(\ref{fracrhobs}) and (\ref{mstar}), $\rho_{V_{\rm end}}$
and $\rho_{m_{\rm end}}$ [Eqs.(\ref{rhoend}) and (\ref{rhochi})
respectively], $t_{\rm end}$ [Eq.(\ref{tend})], and $t_{\rm dis}$
[Eq.(\ref{tdis})], we obtain
\begin{equation}
\gamma\cong \frac{6.7\times 10^{-3}
F^{-5/4}}{H_0/M_{Pl}}\,.\label{gamma62}
\end{equation}

From Eqs.(\ref{gamma62}), (\ref{gdelta}), and (\ref{deltah}) for the
SU(5) model, the initial value of the Hubble parameter $H_{\rm in}$
of the inflation era for $F=0.5$ and $0.3$ is
\begin{equation}
H_{\rm in}\simeq H_0 \simeq M_{Pl}\,. \label{Hinit}
\end{equation}

\section{Reheating of the Universe}

In this section, the reheating of the Universe in our model is
discussed. It is based on the discussion of this epoch in
\cite{partprod2}.

The particles $m$ have a decay rate into relativistic particles,

\begin{equation}
\Gamma_{mr}= f\, \Gamma_{Vm} \,,
\end{equation}
where we assume that
\begin{equation}
f \ll 1\,.
\end{equation}

\begin{figure*}
\resizebox{\columnwidth}{!}{\includegraphics{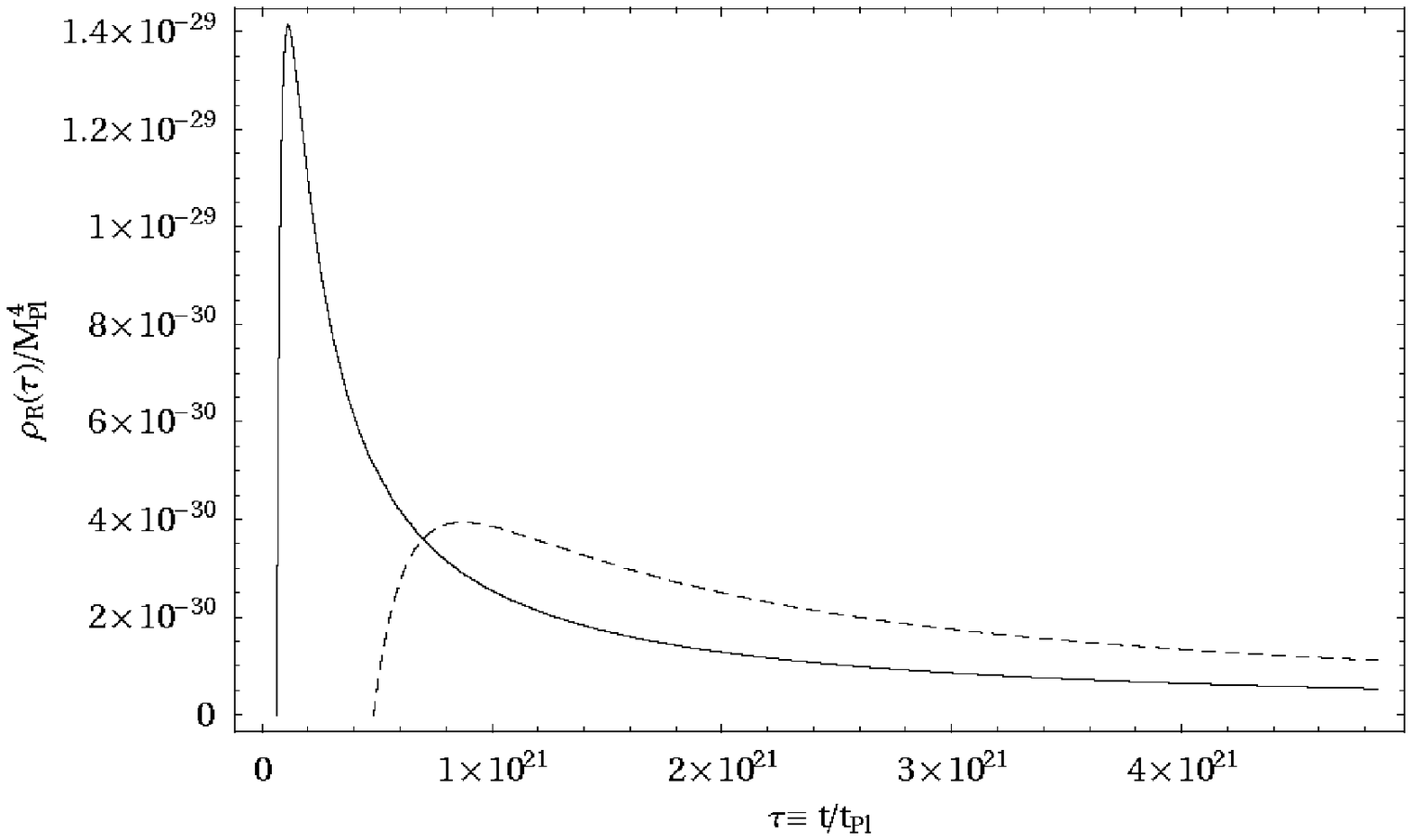}}
\resizebox{\columnwidth}{!}{\includegraphics{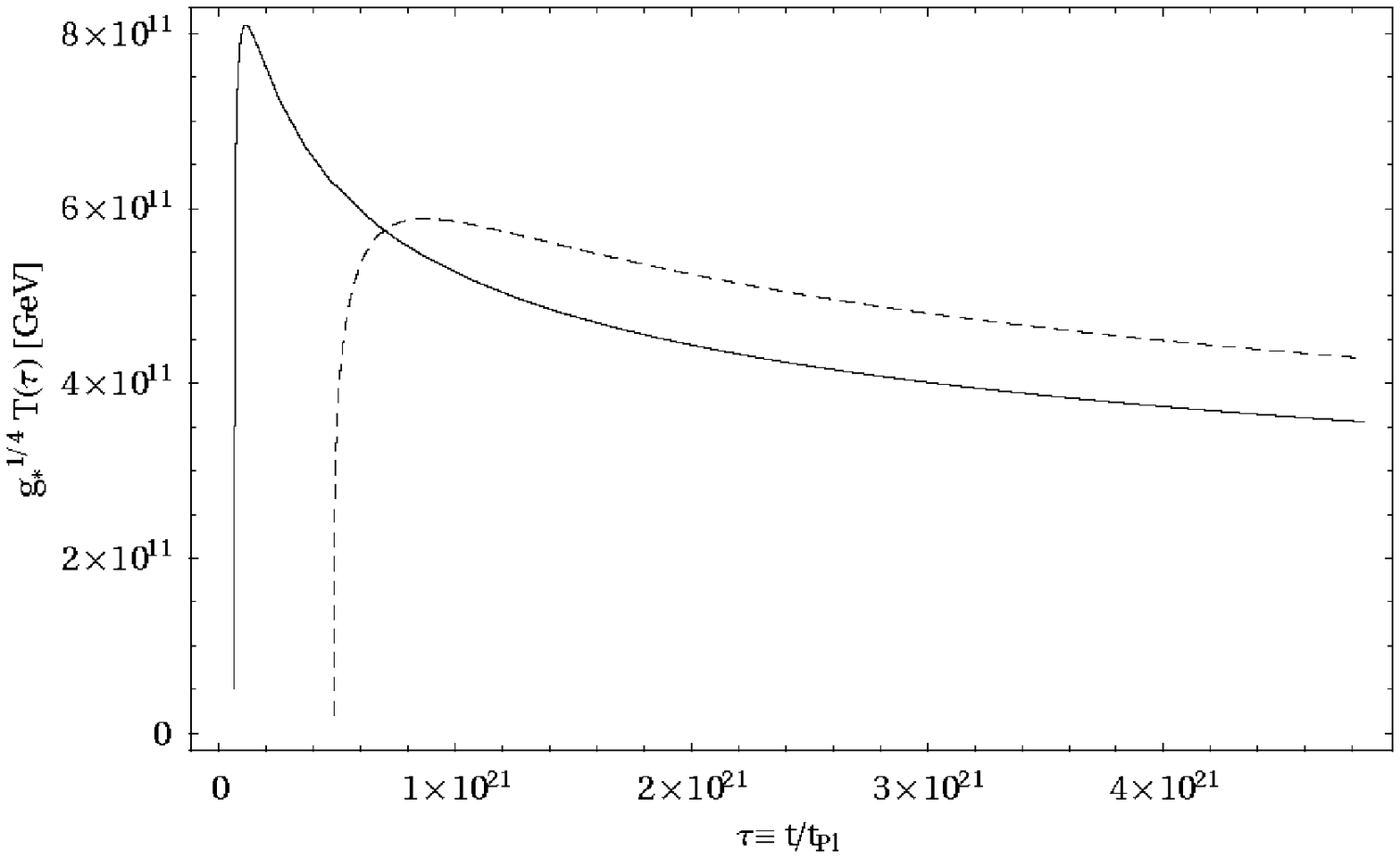}} \caption{a)
Top: The evolution of the radiation energy density $\rho_r$
[Eq.(\ref{rhorr})] (in units of $M_{Pl}^4$) as a function of time
(in units of $t_{\rm Pl}$) for $f=10^{-5}$, starting from $t_{mri}$
(when $\rho_r\simeq 0$) for $F=0.5$ (solid curve) and $F=0.3$
(dashed curve). b) Botton: The temperature (in units of
$g_{\ast}^{-1/4}$GeV) dependence on time [Eqs.(\ref{gstar}) and
(\ref{rhorr})] during the decay of the $m$ particles into
relativistic particles. } \label{rhotemp}
\end{figure*}

The equation describing the decay of the $m$ particles is
\begin{equation}
\dot{\rho}_m + 3 H \rho_m =  - \rho_m \Gamma_{mr}\,,
\end{equation}
which has the analytical solution,
\begin{equation}
\rho_m = \rho_{mi} \left(\frac{a_I}{a}\right)^3 e^{-\Gamma_{mr}
(t-t_{mri} )}\,,
\end{equation}
where $a_I$ is the cosmic scale factor at $t=t_{mri}$. For the
radiation energy density, $\rho_r$, the evolution equation is
\begin{equation}
\dot{\rho}_r + 4 H \rho_r =  \rho_m \Gamma_{mr}\,, \label{rhorreq}
\end{equation}
where $\rho_r$ is the energy density of the relativistic decay
products. In order to obtain the energy density of the relativistic
particles as a function of time, we also need the Friedmann
equation,
\begin{equation}
H^2= \frac{8\pi G}{3}\, (\rho_m + \rho_r)\,.\label{Handrho}
\end{equation}

From $t=t_{mri}(\cong t_{\rm dis})$ until $t=t_{RH}(\equiv
\Gamma_{mr}^{-1})$, the $m$ particles dominate the mass density and
the Universe is matter dominated, $a(t)\propto t^{2/3}$, with
$\rho_r\sim 0$ at $t_{mri}$. During the $m$-dominated epoch, an
approximate solution for $\rho_r$ is given by
$$
\rho_r(t)\simeq \,\frac{\rho_{mi}\,(\Gamma_{mr} t_{mri}^2)}{10\pi
t}\left[1-\left( \frac{t}{t_{mri}}\right)^{-5/3} \right] $$
\begin{equation}
\simeq\,\frac{\sqrt{(6/\pi)}}{10} \rho_{mi}\,(\Gamma_{mr} t_{mri})
\left(\frac{a}{a_I} \right)^{-{3}/{2}} \left[1-\left(
\frac{a}{a_I}\right)^{-{5}/{2}} \right] \, \label{rhorr}
\end{equation}
\cite{KolbTurner}. Thus, $\rho_r$ rapidly increases from $\simeq 0$
at $t_{mri}$ to a value of $\simeq \rho_{mi}\,(\Gamma_{mr} t_{mri})$
at $t_{RH}$, decreasing thereafter as $a^{-3/2}$.

\begin{table*}
\caption{\label{table3}The maximum $T_{\rm max}$ and reheating
$T_{RH}$ temperatures (in units of $g_{\ast}^{1/4}$GeV), the times
$t_{mri}$, $t_{\rm max}$, and $t_{RH}$, for $f=10^{-5}$ and
$10^{-10}$ and $F=0.5$ and 0.3, where $g_{\ast}$ is the number of
relativistic degrees of freedom.}
\begin{ruledtabular}
\begin{tabular}{lllllll}
$F$   &   $f$   &   $T_{\rm max}\,[g_{\ast}^{1/4}{\rm GeV}]$ &
$T_{RH}\,[g_{\ast}^{1/4}{\rm GeV}]$ & $t_{mri}\,[t_{Pl}]$ &
$t_{\rm max}\,[t_{Pl}]$  &  $t_{RH}\,[t_{Pl}]$  \\
\hline
$0.5$ & $10^{-5}$ & $\simeq 1.14\times 10^{12}$ & $\simeq
5.93\times 10^{10}$ & $\simeq 6.28\times 10^{19}$ &
$\simeq 1.13\times 10^{20}$ & $\simeq 4.85 \times 10^{25}$\\
$0.5$ & $10^{-10}$ & $\simeq 6.39\times 10^{10}$ & $\simeq
1.39\times 10^{8}$ & $\simeq 6.28\times 10^{19}$ & $\simeq
1.13\times 10^{20}$ & $\simeq 4.85 \times 10^{25}$\\
$0.3$ & $10^{-5}$ & $\simeq 8.26\times 10^{11}$ & $\simeq
3.19\times 10^{10}$ & $\simeq 4.85\times 10^{20}$ &
$\simeq 8.74 \times 10^{20}$ & $\simeq 6.28 \times 10^{24}$\\
$0.3$ & $10^{-10}$ & $\simeq 4.64\times 10^{10}$ & $\simeq
1.01\times 10^{8}$ & $\simeq 4.85\times 10^{20}$ &
$\simeq 8.74 \times 10^{20}$ & $\simeq 6.28 \times 10^{24}$\\
\end{tabular}
\end{ruledtabular}
\end{table*}

Once the relativistic decay products interact sufficiently, they
thermalize and we have
\begin{equation}
\rho_r=g_{\ast} \pi^2 T^4/30\,,\label{gstar}
\end{equation}
where $g_{\ast}$ is the number of relativistic degrees of freedom
and is generally estimated to be $100\lesssim g_{\ast} \lesssim
1000$. What is commonly called the reheat temperature, $T_{RH}$, is
not the maximum temperature of the Universe, $T_{\rm max}$, which is
given by
\begin{equation}
T_{\rm max} \cong 0.8\,(\,\rho_{mi}\Gamma_{mr} t_{mri})^{1/4}\,
g_{\ast}^{-1/4} \,. \label{tempmax}
\end{equation}
The reheat temperature at the beginning of the radiation dominated
epoch, $T_{RH} \equiv T(t_{RH}={\Gamma_{mr}^{-1}})$, is given by
\begin{equation}
T_{RH}\cong 0.55\, (\,\rho_{mi}^{1/2}\,\Gamma_{mr}
t_{mri})^{1/2}\,g_{\ast}^{-1/4} \,\label{tempreh}
\end{equation}
and the ratio of $T_{\rm max}$ to $T_{RH}$ is
\begin{equation}
\frac{T_{\rm max}}{T_{RH}}=
\sqrt[4]{\frac{\Gamma_{Vm}}{\Gamma_{mr}}}\cong
\frac{1.45}{f^{1/4}}
\end{equation}
since $\Gamma_{mr}=f\,\Gamma_{Vm}$, where $f\ll 1$. In
Table~\ref{table3}, we show the values of $T_{\rm max}$
[Eq.(\ref{tempmax})] and $T_{RH}$ [Eq.(\ref{tempreh})] for
$r=6.8\times 10^{-4}$, $f=10^{-5}$ and $10^{-10}$, and $F\equiv
m/M=0.5$ and 0.3.

In Fig.~1, we show $\rho_r$ [Eq.(\ref{rhorr})] and the temperature
as a function of time [Eq.(\ref{gstar})] for $f=10^{-5}$ and
$F=0.5$ and 0.3 during the time interval $t_{mri}<t< t_{RH}$. The
times $t_{mri}$, $t_{RH}$, and $t_{\rm max}$ are shown in
Table~\ref{table3}.


\section{Conclusions and Discussions}

We presented here a model which relates the Starobinsky and
$\epsilon R^2$ models, both of which predict inflation, to the
reheating era by a massive conformally coupled particle. In the
original Starobinsky model, the coupling to the reheating era was
made by massless non-conformally coupled particles. Here we assumed
that non-conformally coupling to gravitation does not exist.

In the Starobinsky model, inflation is due to vacuum fluctuations.
Inflation is due to a modification of the gravitational Lagrangian
in the $\epsilon R^2$ model. In both models, inflation takes place
without the need for a scalar inflaton field.

The end of the inflation period predicted by the Starobinsky and
$\epsilon R^2$ models is characterized by a parameter $M$
($M=M_{Pl}/\sqrt{48\pi k_1}$ in the Starobinsky model, where $k_1$
is the coefficient of the $^{(1)}H$ tensor
(Eqs.\ref{anomaly},\ref{Htensor}) of the expectation value of the
vacuum energy-momentum tensor due to vacuum fluctuations and
$M=M_{Pl} /\sqrt{24}\epsilon$ in the $\epsilon R^2$ model).

In our model, we assumed the existence of particles of mass $m$ and
linked their conformal gravitational production during inflation to
the observed scale invariant density fluctuations. Their
gravitational production at the end of inflation was linked to the
matter density of the Universe.

Our model of two massive scalar particles $M$ and $m$ has a certain
resemblance to the double inflation model of Gottl\"ober, M\"uller,
and Starobinsky  \cite{2infla91}. Gottl\"ober, M\"ucket and
Starobinsky discussed the confrontation of the model with
observations \cite{2infla94}. In the simple chaotic inflation model,
the scalar potential, $(1/2) m^2 \phi^2$, is characterized by the
mass $m$. The gravitational term, $R^2/M^2$, is characterized by the
mass $M$ in the late inflation in the Starobinsky model. In the
double inflation model of Gottl\"ober, M\"uller and Starobinsky,
both the scalar field potential, $({1}/{2}) m^2 \phi^2$, and the
gravitational term, $R^2/M^2$, are present. Thus this double
inflation model is described by two masses $M$ and $m$. It is to be
noted that in our model, described by the two masses $M$ and $m$,
only the mass $M$ describes the inflation era, whereas the mass $m$
is a free particle, produced at the end of inflation, linking the
inflation era to the reheating era. The $m$ particles in our model,
have a decay time into relativistic particles $t_{\rm RH}$ very much
greater than $t_{\rm dis}$, the time for the production of the $m$
particles at the end of inflation.

Our model depends upon the parameter $H_0$, the normalization of the
Starobinsky inflation solution for the Hubble parameter near the
Planck era [Eq.(\ref{Hsol})]. The parameter $H_0$ is on the order of
$M_{Pl}$ and depends on $N_0,\,N_{1/2},$ and $N_1$, the number of
quantum matter fields in the vacuum of spin 0, 1/2, and 1,
respectively. To evaluate $H_0$, we assume a minimal SU(5) particle
content and obtain $H_0\cong 0.74\,M_{Pl}$ [Eq.(\ref{H0su5})]. The
time at the end of inflation depends on $H_0$ and $M$. For $r\simeq
6.8\times 10^{-4}$ and $M\simeq 1.15\times 10^{-6}\,M_{\rm Pl}\simeq
1.4\times 10^{13}\,{\rm GeV}$, as predicted by the Starobinsky model
for $F=0.3$, for example, we find $t_{\rm end}\simeq 1.37\times
10^{11}\, t_{Pl} \approx 10^{-32}\,{\rm sec}$ and $t_{\rm dis}\simeq
4.8\times 10^{20}\,t_{Pl}\approx 10^{-23}\,{\rm sec}$, respectively.

In the future, the particle $m$ could be incorporated into a
particle physics theory that would define the decay time into
relativistic particles $t_{RH}$, the reheating time. From $f\equiv
t_{\rm dis}/t_{RH}$, we evaluate the maximum temperature of the
Universe $T_{\rm max}$ and the reheat temperature $T_{RH}$ as a
function of $g_{\ast}$ (the number of degrees of freedom of the
relativistic particles) and $F\equiv m/M$. The times $t_{Mri}$,
$t_{\rm max}$, and $t_{RH}$ can, then, also be evaluated. A measured
value of $r$, the ratio of tensor to scalar fluctuations, that is
appreciably different from $r=6.8\times 10^{-4}$ [Eq.(72)] would
discard our model (as well as the Starobinsky and $\epsilon R^2$
models).

In order not to overproduce gravitinos, it is frequently suggested
that $T_{RH}\lesssim  10^{9}\,{\rm GeV}$  \cite{gravitino}. If
this is, indeed,  a true upper limit for $T_{RH}$, it puts limits
on the possible values for $f$ and $F$ from the above analysis.
However, it is to be noted that, for our model, $10^{9}\,{\rm
GeV}$ is not a strong upper limit for $T_{RH}$ since supersymmetry
is still not a well developed theory.

\noindent {\bf Acknowledgments.} R.O. thanks the Brazilian agencies FAPESP (grant
00/06770-2) and CNPq (grant 300414/82-0) for partial support. A.P. thanks FAPESP for
financial support (grants 03/04516-0 and 00/06770-2).



\end{document}